\documentclass[aps,pre,twocolumn,showpacs,groupedaddress]{revtex4}

\usepackage{graphicx}

\begin{document}


\title{Propagation dynamics on the Fermi-Pasta-Ulam lattices}



\author{Zongqiang Yuan}
\affiliation{Department of Physics and the Beijing-Hong Kong-Singapore Joint Centre for Nonlinear and Complex Systems (Beijing), Beijing Normal University, Beijing 100875, China}
\author{Zhigang Zheng}
\email[]{zgzheng@bnu.edu.cn}
\affiliation{Department of Physics and the Beijing-Hong Kong-Singapore Joint Centre for Nonlinear and Complex Systems (Beijing), Beijing Normal University, Beijing 100875, China}



\begin{abstract}
The spatiotemporal propagation of a momentum excitation on the finite Fermi-Pasta-Ulam lattices is investigated. The competition between the solitary wave and phonons gives rise to interesting propagation behaviors. For a moderate nonlinearity, the initially excited pulse may propagate coherently along the lattice for a long time in a solitary wave manner accompanied by phonon tails. The lifetime of the long-transient propagation state exhibits a sensitivity to the nonlinear parameter. The solitary wave decays exponentially during the final loss of stability, and the decay rate varying with the nonlinear parameter exhibits two different scaling laws. This decay is found to be related to the largest Lyapunov exponent of the corresponding Hamiltonian system, which manifests a transition from weak to strong chaos. The mean-free-path of the solitary waves is estimated in the strong chaos regime, which may be helpful to understand the origin of anomalous conductivity in the Fermi-Pasta-Ulam lattice.
\end{abstract}


\pacs{05.45.Yv, 44.10.+i}


\maketitle

\section{Introduction}

The study of transport process of matter and energy is of fundamental importance in understanding numerous nonequilibrium phenomena occuring in nature. Heat conduction is one of the most important manners of energy transport. Recently, heat conduction in low-dimensional materials has attracted much attention among physicists for the reason that classical one-dimensional lattices frequently exhibit anomalous heat conduction behavior, \textit{i.e.}, the thermal conductivity depends crucially on the size of the material~\cite{Lepri1997,Aoki2001,Lepri2003,Dhar2008c,Li2012}. This arouse a tide of interest in the microscopic foundation of normal heat conduction, and a number of viewpoints on the relation between heat conduction and dynamical properties have been proposed, such as chaos, mixing, energy diffusion, and so on.

The manipulation of heat flow is an important and practical issue, which has been developed rapidly in recent years~\cite{Terraneo2002,Li2004,Chang2007a,Wang2007,Wang2008a,Wang2008}. Thermal rectifier had been experimentally realized in nanoscale systems~\cite{Chang2006a,Kobayashi2009}. By periodically modulating thermal baths, heat flow can even be created and controlled at strict zero thermal bias~\cite{Ren2010}. A relevant problem is the competition between time scale of the manipulation of the thermal bath and the relaxation time scale of the heat flow along the lattice. Therefore it is important to study the propagation process of energy in nonlinear low-dimensional systems from the microscopic point of view.

In studies of heat conductions of nonlinear lattices, the thermal baths usually contact with the system by coupling the particles at two ends. The influence of the thermal bath on the lattice can be considered as a series of stochastic perturbations. These perturbations start from the ends of the lattice and propagate along the lattice with a finite speed. It is a significant topic to study the propagation dynamics and dispersion behaviors of energy pulses on the lattice. Therefore, we may explore the evolution behavior of a single pulse excited at one end of the lattice as the first step. This may give us a more profound microscopic understanding of the transfer process of heat on low-dimensional nonlinear lattices. In this aspect, energy propagation for an excitation of a single particle on infinite nonlinear lattices has been discussed~\cite{Zavt1993,Sarmiento1999,Rosas2004}. Practically, systems have finite sizes, and the propagation of energy on materials usually possesses a finite time scale. Therefore, the influence of finite length of low-dimensional materials should be taken into account.

In the present paper, we investigate the propagation behavior of an energy pulse initially excited at one end of a finite Fermi-Pasta-Ulam $\beta$ (FPU-$\beta$) lattice. The initially excited pulse may propagate along the lattice in a solitary wave manner accompanied by phonon tails. For a moderate nonlinearity, the solitary wave can propagate coherently for a long time before its collapse, which is called the long-transient propagation state. The lifetime of the long-transient propagation state displays a sensitive dependence on the nonlinear parameter $\beta$. The energy of the solitary wave decreases exponentially during the collapse process, which is irrelevant to the boundary conditions. The decay rate against $\beta$ exhibits two different scaling laws which is found to be related to the largest Lyapunov exponent of the corresponding Hamiltonian system.~The multiple-peak structure of the lifetime of the long-transient propagation state on $\beta$ is understood by the residual high-dimensional Kolmogorov-Arnold-Morse (KAM) tori of the Hamiltonian lattice system in the parameter regime of $\beta$ with moderate stochasticity. Our results presented in this paper may be helpful to understand miscellaneous recently studied macroscopic heat phenomena based on microscopic energy wave properties on the nonlinear lattices.

\section{The Fermi-Pasta-Ulam model}

The famous FPU model was initially introduced by Fermi, Pasta, and Ulam to investigate the energy equipartition problem and the ergodic hypothesis in nonlinear systems~\cite{Fermi1955}. The attempt to resolve the mystery of the FPU recurrence has led to the discovery of solitons~\cite{Zabusky1965}. Later tremendous progresses on FPU model have been extended to studies on intrinsic localized modes in perfect lattices, Bose-Einstein condensates, stochastic resonance, and so on~\cite{Sievers1988,Flach2005,Villain2000,Miloshevich2009}. Recently the FPU lattice was studied in relating to heat conductions in low-dimensional systems~\cite{Li2005a,Mai2007,Dhar2008b,Li2010}. In this paper we adopt the FPU-$\beta$ model as our prototype to study the energy transport on nonlinear lattices. The Hamiltonian of the FPU-$\beta$ lattice consisting of $N$ particles with open boundary condition can be written as
\begin{eqnarray}
\label{DefineHamitonian}
H &=& \sum _{i=1}^N \frac{p_i^2}{2} + \sum _{i=1}^{N-1} V(q_{i+1},q_i), \nonumber\\
V(q_{i+1},q_i)&=& \frac{k}{2}\Bigl(q_{i+1}-q_i\Bigr)^2 + \frac{\beta}{4}\Bigl(q_{i+1}-q_i \Bigr)^4,
\end{eqnarray}
where $p_i$ and $q_i$ denote the momentum and the displacement from the equilibrium position of the $i$-th particle, respectively. The local energy of the \textit{i}-th particle can be defined as $E_i=\frac{p_i^2}{2}+\frac{1}{2}V(q_{i+1},q_i)+\frac{1}{2}V(q_i,q_{i-1})$. In the absence of the quartic term, \textit{i.e.}, $\beta=0$, the above Hamiltonian reduces to a one-dimensional harmonic chain, which is integrable and can be analytically solved. The presence of the anharmonic terms breaks the integrability and brings forth the intermingling of regular and chaotic motions in phase space~\cite{Antonopoulos2006}.

To explore the evolution behavior of a single pulse excited at one end of the lattice, we may impart an initial momentum excitation to the first particle of an initially quiescent lattice. In our numerical simulation, we adopt the fourth-order symplectic method in order to solve the dynamics of the FPU lattice as a Hamilton dynamical system. We further fix the harmonic coefficient $k=0.5$, particles of the lattice $N=50$ and energy of the initial excitation $E=50$ (this also gives the total energy of the lattice) throughout the simulation.

\section{The long transient propagation}

\begin{figure}
\includegraphics[width=\linewidth]{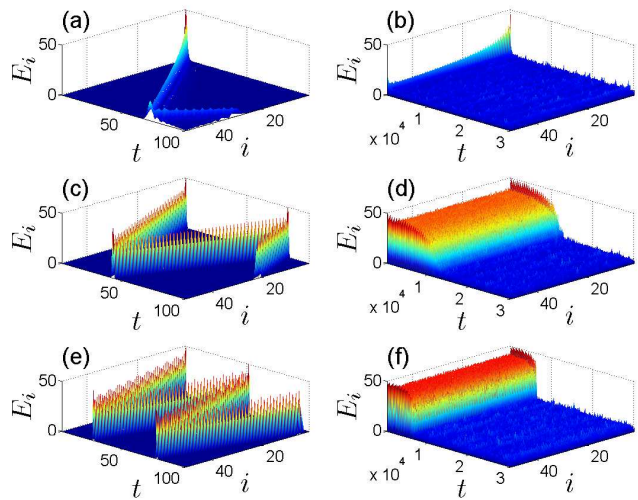}
\caption{(Color on-line)~Spatiotemporal propagation behavior of the initial momentum excitation imposed on the first particle of the lattice. (a), (b) $\beta=0.001$, (c), (d) $\beta=0.05$, and (e), (f) $\beta=0.4$. The left and right columns correspond to the short and long time scales, respectively.}
\label{Evolution}
\end{figure}

Here we are concerned with the destiny of an initial pulse on a lattice with finite size. For a weak nonlinearity $\beta$, the dispersion effect dominates this weakly nonlinear system and leads to the rapid collapsing of the initial local excitation, as shown in Fig.~\ref{Evolution}(a) for $\beta=0.001$. When we increase $\beta$, the initial excitation may excite a solitary wave. This is shown in Figs.~\ref{Evolution}(c) for $\beta=0.05$ and~\ref{Evolution}(e) for $\beta=0.4$. These are consistent with previous works~\cite{Wattis1993,Friesecke1994,Zhang2000,Zhang2001}. The solitary wave will propagate with no decay if the lattice is extended to infinity. However, with finite lattice, one finds two distinctively different stages as depicted in Figs.~\ref{Evolution}(d) and~\ref{Evolution}(f). In the first stage, the energy pulse initiated at the first particle can be coherently transferred to its neighboring and other particles, and this pulse forms a solitary wave along the lattice. The propagation of the solitary wave keeps stable for a long time. The second stage comes when the solitary wave loses its stability and collapses in a rather short duration, and the energy of the solitary wave is distributed to all particles in the lattice. This behavior is very interesting, indicating that the solitary wave can dominate by suppressing the phonon waves for a long time.

\begin{figure}
\includegraphics[width=\linewidth]{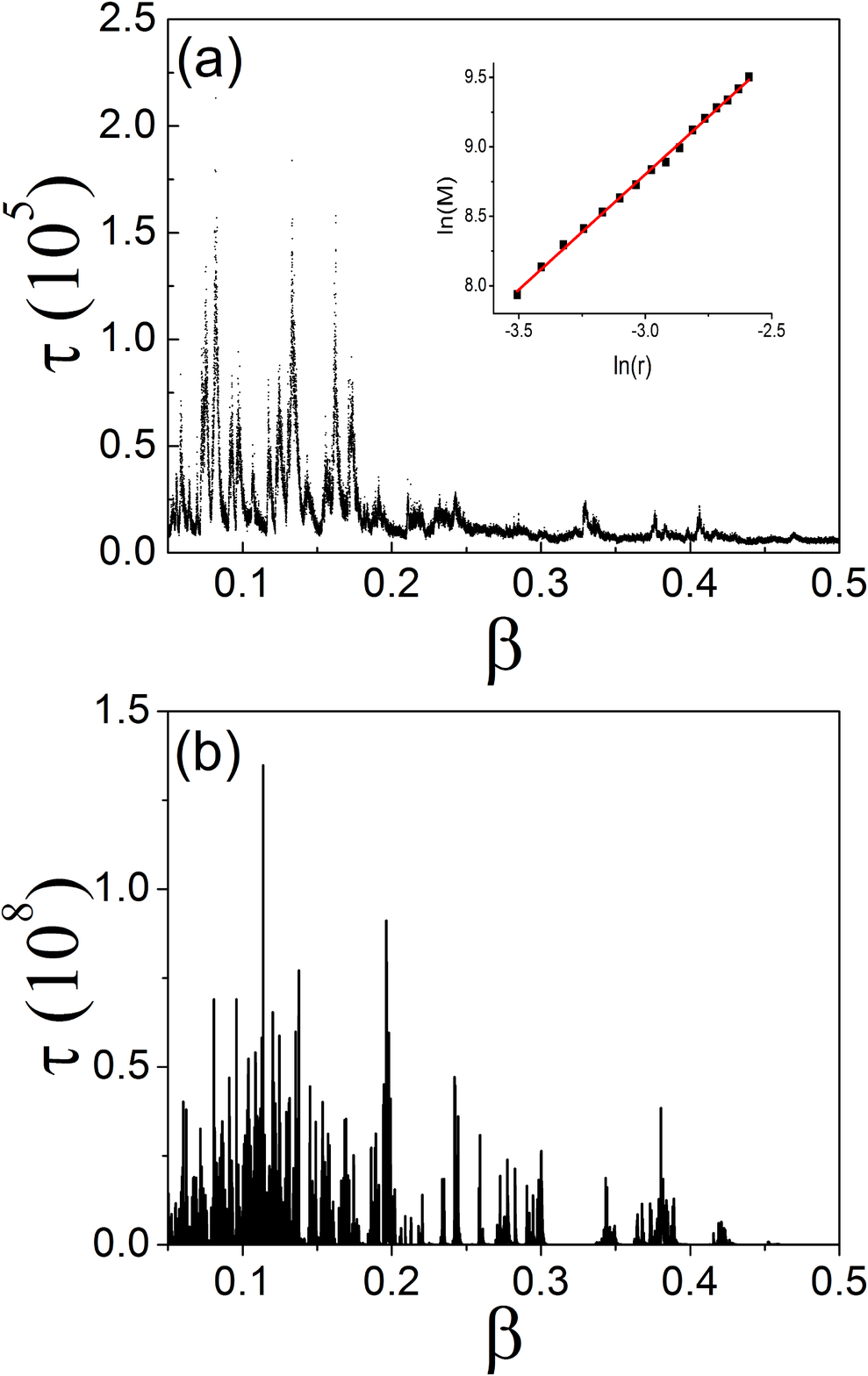}
\caption{(Color on-line)~The lifetime $\tau$ of the solitary wave. (a) and (b) corresponds to the cases of open and periodic boundary conditions, respectively. To numerically get the fractal dimension of the hierarchical multiple-peak structure for the case of open boundary condition, result of a typical point count (see text) is shown in the inset of (a).}
\label{Lifetime}
\end{figure}

The lifetime $\tau$ of the solitary wave (the long-transient propagation state) as the function of the anharmonic parameter $\beta$ for the case of open boundary condition is shown in Fig.~\ref{Lifetime}(a). The value of $\tau$ is sensitive to $\beta$, especially for $\beta \in (0.05, 0.2)$ where a hierarchical multiple-peak structure is found. By adopting the method proposed in ref.~\cite{Forrest1979}, the fractal dimension of this hierarchical structure can be estimated as follows. The lifetime $\tau$ is re-scaled and a center point on the image is picked at random, and then a series of nested circles of different sizes are placed around it and the number of points (lifetime data for $\beta$) in each circle counted. The number of points $M$ in a circle with a radius $r$ satisfies as $M \propto r^D$, where $D$ is the dimension of the measured object. We numerically get the fractal dimension to be $D \approx 1.66$. This result implies the complex dynamical stability of the transient propagation by varying the nonlinear parameter. As will be shown below, this property indicates a connection between the stability of solitary wave and the structure of the phase space of the Hamiltonian lattice system.

\section{Phonon-soliton interaction}

\begin{figure}
\includegraphics[width=\linewidth]{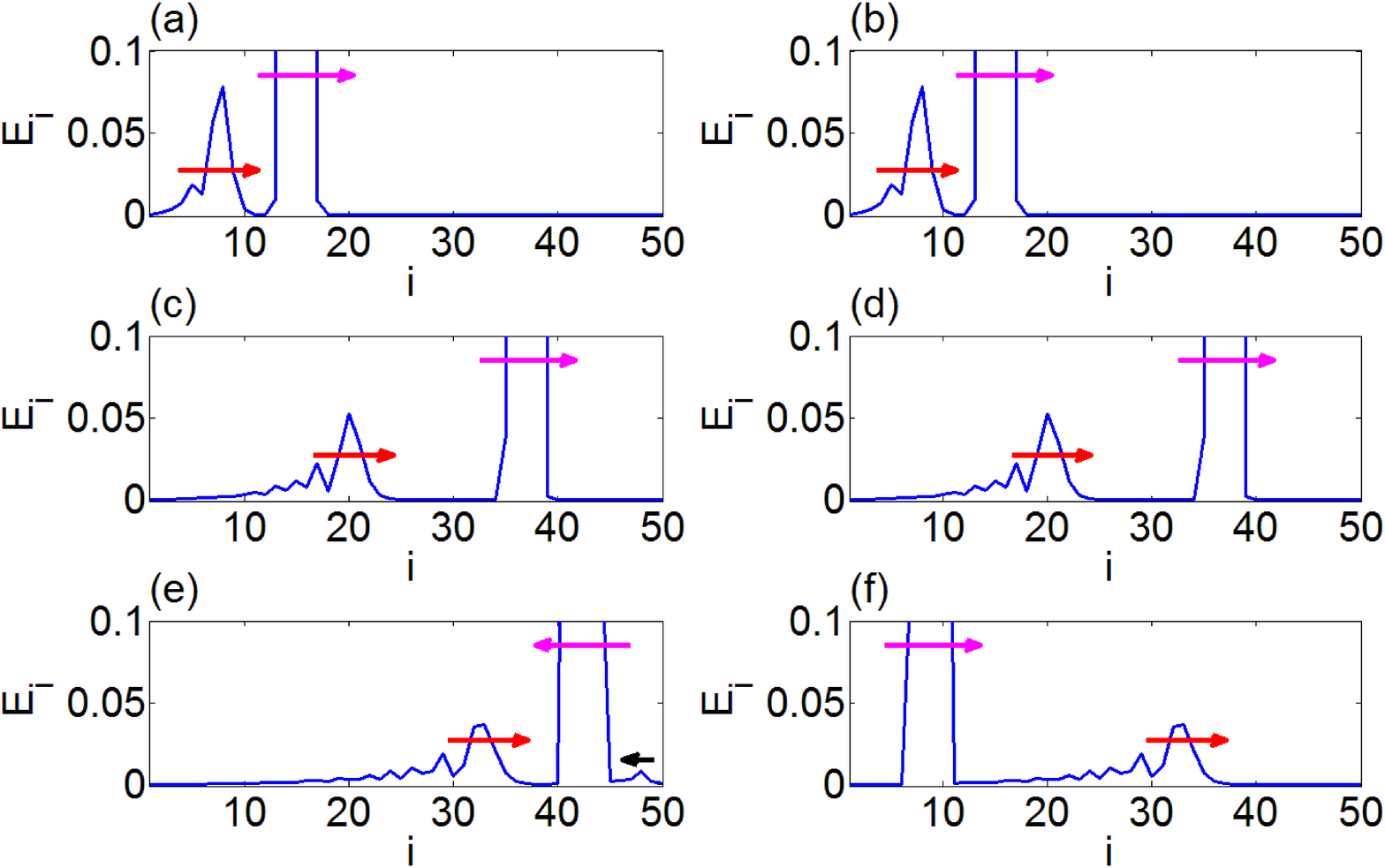}
\caption{(Color on-line)~Energy distribution profiles among the particles for the FPU-$\beta$ lattice with $\beta=0.05$ for different moments: (a), (b) $t=12$, (c), (d) $t=30$, and (e), (f) $t=48$. The left and right columns correspond to the cases of open and periodic boundary conditions, respectively. Because the first peak is much higher than others, we only plot the energy profile in (-0.001,0.1) to get a clearer observation of the wave tails.}
\label{EnergyProfile}
\end{figure}

It is important to study the competition of various propagation modes (waves) on the lattice to understand the above results. A sufficiently large momentum excitation initially imposed on the first particle excites not only a solitary wave, but also a small-amplitude tail. The solitary wave moves along the lattice much faster than the phonon-wave tail, which disperses due to the dispersion property of the phonon modes. Moreover, the reflections of the solitary wave at both ends of the lattice can also excite additional small-amplitude tails for the case of open boundary condition. In Figs.~\ref{EnergyProfile}(a), (c) and (e), the snapshots of propagations of the energy waves along the lattice for the open boundary condition case are plotted for different moments. It is clear that the solitary wave moves with a hierarchy of lower pulses that move slower, and the heights of these small pulses decrease during their motion along the lattice. As the solitary wave moves to the boundary, it will be bounced back with a radiation of additional phonon waves [Fig.~\ref{EnergyProfile}(e)]. The multiple collisions between the solitary wave and phonon waves give rise to the instability of the solitary wave and its collapse.

This phonon-soliton-interaction mechanism is also valid for other types of boundary conditions, \textit{e.g.}, the periodic boundary condition. Technically a single solitary wave can be produced by initially exciting an energy pulse at one end of an open chain and then connecting both ends when the solitary wave arrives at the middle of the lattice. Different from the open boundary condition case, for the periodic boundary condition, there is no reflection of the solitary wave at the boundary. Therefore the solitary wave moves unidirectionally, and only the phonon-wave tail due to the initial excitation can be found, as shown in Figs.~\ref{EnergyProfile} (b), (d) and (f). Due to the lack of additional excitations of phonons at the boundary, the lifetime of the solitary wave moving on a circular topology of the lattice should be much longer than that on an open lattice, as shown in Fig.~\ref{Lifetime}(b). However, because the solitary wave moves faster than its initial phonon tail, they will frequently collide when they meet. This interaction will eventually lead to the collapse of the solitary wave.

\section{Decay process of the solitary wave}

By resorting to the evolution of the energy of the small-amplitude tails, we now focus on the collapse process of the solitary wave due to the interaction with phonons mentioned above. For the lattice system we are studying here, the energy of the tails is defined as the residual energy of the solitary wave. Due to the spatial localization of the solitary wave, one can write the tail energy as
\begin{equation}
\label{DefineEtails}
 E_{tails} = E - \sum_{i=i_c-i_n}^{i_c+i_n} E_i(t),
\end{equation}
where $E$ is the total energy of the system, $E_i$ is the local energy of the \textit{i}-th particle, $i_c(t)$ is the center position of the solitary wave at time $t$, and $i_n$ denotes the number of the left/right neighboring particles of the center particle of the solitary wave packet. Numerically $i_n=2$ is enough due to the energy localization of the solitary wave. The increase of $E_{tails}$ corresponds to the dissipation of the solitary wave energy considering that the total energy of the system is conserved.

\begin{figure}
\includegraphics[width=\linewidth]{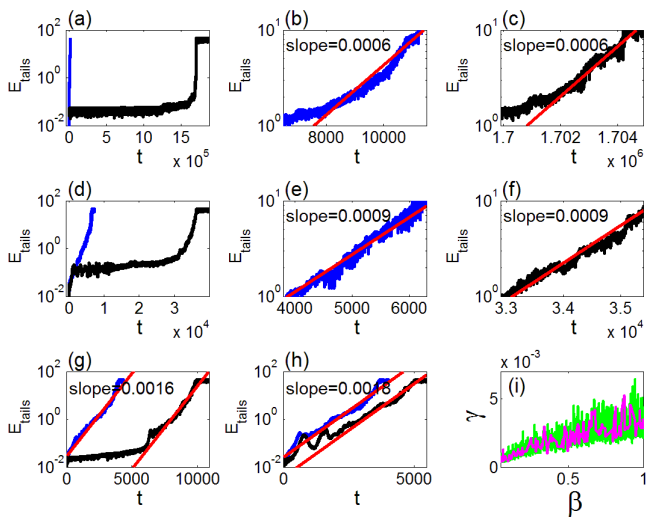}
\caption{(Color on-line)~Energy increase of the tails with open (blue line) and periodic (black line) boundary conditions: (a-c) $\beta=0.2$, (d-f) $\beta=0.4$, (g) $\beta=0.56$, (h) $\beta=0.72$. (b), (c) and (e), (f) enlarge the final stages of the collapse processes in (a) and (d), respectively. The red lines are for guiding the eyes. (i) Decay rate $\gamma$ against $\beta$ for the cases of open (green line) and periodic (pink line) boundary conditions.}
\label{Collapse}
\end{figure}

We present the evolution of $E_{tails}$ for several typical values of $\beta$ in Figs.~\ref{Collapse}(a)-(h). Although the lifetime of the solitary wave varies for different boundary conditions and different $\beta$, the collapse process of the solitary wave exhibits the same scenario, \textit{i.e.}, the energy of the solitary wave decays exponentially during the final loss of stability. We label the exponential decay rate by $\gamma$. In Fig.~\ref{Collapse}(i), the decay rate $\gamma$ against $\beta$ is given for both open and periodic boundary conditions. The consistency of different types of boundary conditions indicates that the final loss of stability of the solitary wave is irrelevant to the boundary conditions.

\begin{figure}
\includegraphics[width=\linewidth]{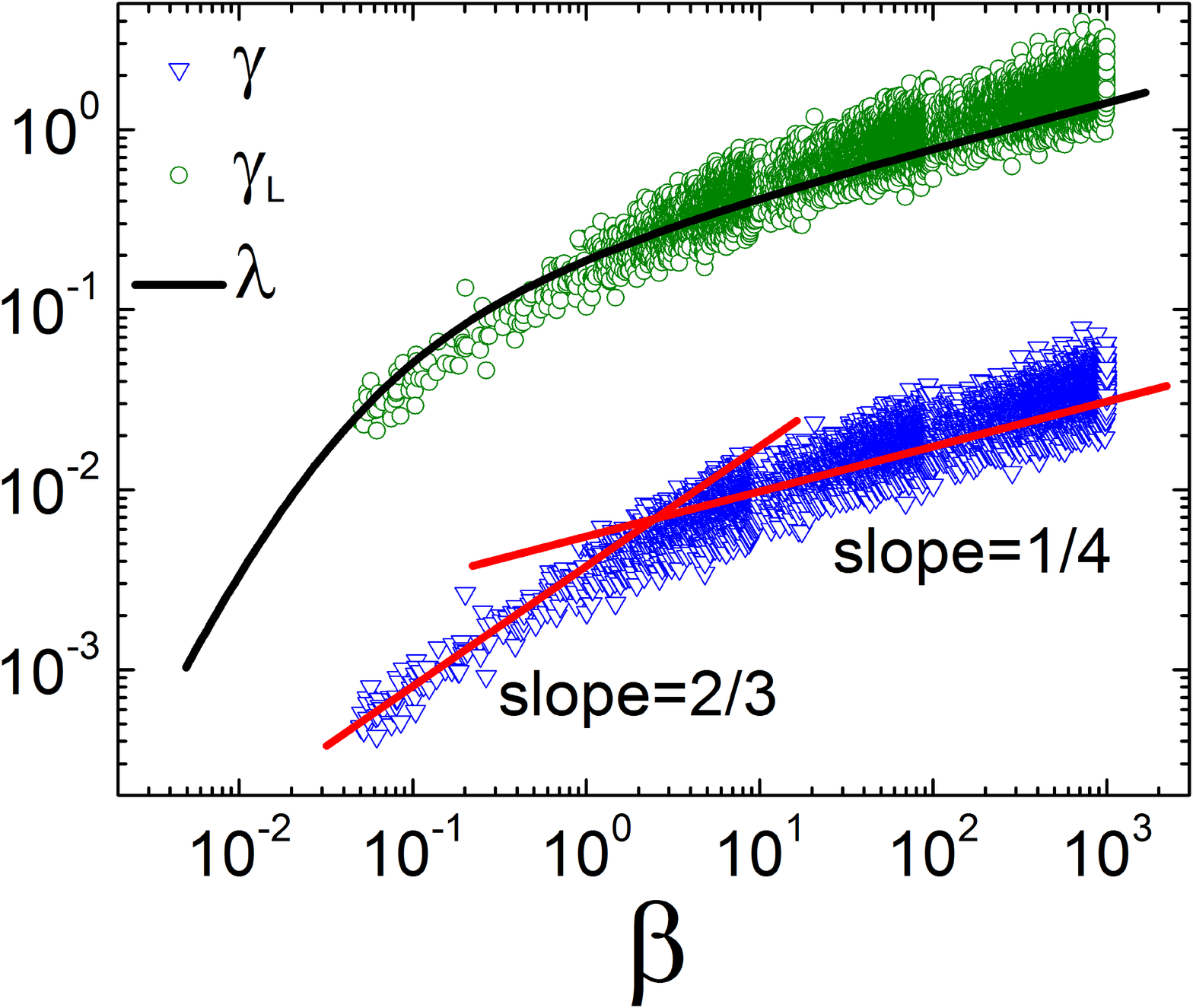}
\caption{(Color on-line)~Decay rate $\gamma$ against the nonlinear parameter $\beta$ for the case of open boundary condition (blue triangles). Green circles correspond to the rescaled decay rate $\gamma_L$. The black line corresponds to the largest Lyapunov exponent $\lambda$ computed according to the analytic expression (\ref{formula for lamda}). The red lines are for guiding the eyes.}
\label{DecayExponents}
\end{figure}

In Fig.~\ref{DecayExponents}, the decay rate $\gamma$ is computed numerically in a larger scale of the nonlinear parameter $\beta$ for the case of open boundary condition. It can be seen that $\gamma$ against $\beta$ displays two different scaling laws $\gamma \propto \beta^{\kappa}$. For lower values of $\beta$, the scaling exponent $\kappa \approx 2/3$. For larger nonlinear parameter $\beta$, $\kappa \approx 1/4$. It is instructive to note that a similar result was obtained in studies of the nonlinear Hamiltonian dynamics of the FPU-$\beta$ lattice~\cite{Gallavotti2008}. It was found that the largest Lyapunov exponent $\lambda$ of the system varying with the energy density $\epsilon=E/N$ exhibits a crossover between two scaling laws: $\lambda \propto \epsilon^2$ at low-energy density, and $\lambda \propto \epsilon^{2/3}$ at larger $\epsilon$ values, reaching on an asymptotic value at large energy of $\lambda \propto \epsilon^{1/4}$. One should note that changing the nonlinear parameter $\beta$ is equivalent to changing the energy density $\epsilon$~\cite{Aoki2001}.

We can analytically estimate the largest Lyapunov exponent $\lambda$ of the corresponding Hamiltonian system as a function of $\beta$ following the theoretical approach of Riemannian differential geometry of Newtonian dynamics~\cite{Casetti1995,Casetti1996}. In the geometric approach to Hamiltonian chaos, the dynamics described by the equations of motion is equivalent to a geodesic flow on a Riemannian manifold. Dynamical instability (chaos) is related to curvature fluctuations of the manifolds and is described by means of the Jacobi-Levi-Civita equation for geodesic spread. The analytic formula for $\lambda$ is
\begin{equation}
\label{formula for lamda}
\lambda = \frac{1}{2} \biggl ( \Lambda - \frac{4 \Omega_0}{3 \Lambda} \biggr),
\end{equation}
\begin{equation}
\Lambda = \biggl [ 2 \sigma_{\Omega}^2 \tau + \sqrt{ \biggl( \frac{4 \Omega_0}{3} \biggr)^3 + (2 \sigma_{\Omega}^2 \tau})^2 \biggr]^{1/3},
\end{equation}
\begin{equation}
 2 \tau = \frac{\pi \sqrt{\Omega_0}}{2 \sqrt{\Omega_0(\Omega_0+\sigma_{\Omega})} + \pi \sigma_{\Omega}},
\end{equation}
where $\Omega_0$ and $\sigma_{\Omega}$ corresponds to the average Ricci curvature and its fluctuation, respectively. For our Hamiltonian (\ref{DefineHamitonian}), the explicit expression for the Ricci curvature $k_R$ is
\begin{equation}
\label{kR}
  k_R = 2k + \frac{6\beta}{N} \sum_{i=1}^N (q_{i+1}-q_i)^2,
\end{equation}
where $k$ and $\beta$ correspond to the harmonic and anharmonic coefficients in the Hamiltonian of the FPU-$\beta$ lattice, respectively. Then the expressions for $\Omega_0$ and $\sigma_{\Omega}$ can be derived as
\begin{equation}
\Omega_0 = 2k +  \frac{3k}{\theta} \frac{D_{-3/2}(\theta)}{D_{-1/2}(\theta)},
\label{compute average of Ricci curvature of FPU beta curvature}
\end{equation}
\begin{equation}
\sigma_{\Omega}^2 = \frac{9 k^2}{\theta^2} \biggr \{ 2 - 2 \theta \frac{ D_{-3/2} (\theta)}{D_{-1/2}(\theta)}  - \biggl[ \frac{ D_{-3/2} (\theta)}{D_{-1/2}(\theta)} \biggr ]^2  \biggr \} +
F(\theta),
\label{compute average fluctuation of Ricci curvature of FPU beta curvature}
\end{equation}
where $D_x$ are parabolic cylinder functions. The results are expressed in terms of the parameter $\theta=k \sqrt{\Theta /2 \beta}$, where $\Theta$ is the inverse temperature introduced by the Gibbsian weight $e^{-\Theta H}$. The additional term $F(\theta)$ is
\begin{equation}
\label{corrective}
  F(\theta) = - \frac{\Theta^2}{c_V(\theta)} \biggl ( \frac{\partial \Omega_0 (\theta)}{\partial \Theta} \biggr)^2,
\end{equation}
where the derivative part is
\begin{equation}
  \frac{\partial \Omega_0}{\partial \Theta} = \frac{-3k^3}{8 \beta \theta^3} \biggr \{2 \theta - 2(\theta^2-1) \frac{D_{-3/2}(\theta)}{ D_{-1/2}^2(\theta)} - \theta \biggl[ \frac{ D_{-3/2} (\theta)}{D_{-1/2}(\theta)} \biggr ]^2 \biggr \},
\end{equation}
and the specific heat per particle $c_V$ is found to be
\begin{equation}
  c_V = \frac{3}{4} + \frac{\theta^2}{8} - \frac{\theta(\theta^2-1)}{8} \frac{D_{-3/2}(\theta)}{ D_{-1/2}^2(\theta)} - \frac{\theta^2}{16} \biggl[ \frac{ D_{-3/2} (\theta)}{D_{-1/2}(\theta)} \biggr ]^2.
\end{equation}
Substituting Eqs. (\ref{compute average of Ricci curvature of FPU beta curvature}) and (\ref{compute average fluctuation of Ricci curvature of FPU beta curvature}) into Eq. (\ref{formula for lamda}) yields an analytic expression of $\lambda$ for the FPU-$\beta$ model (\ref{DefineHamitonian}), valid in the thermodynamic limit $N \to \infty$. A relation between the nonlinear parameter $\beta$ and the parameter $\theta$
\begin{equation}
\beta(\theta)=\frac{k^2}{8 \epsilon} \biggl [ \frac{3}{\theta^2} + \frac{1}{\theta} \frac{D_{-3/2}(\theta)}{D_{-1/2}(\theta)} \biggr ]
\label{compute average fluctuation of Ricci curvature of FPU beta parameter}
\end{equation}
allows one to obtain $\lambda$ as a function of $\beta$. We present $\lambda$ against $\beta$ in Fig.~\ref{DecayExponents}. It is clear that the rescaled decay rate $\gamma_L=L\gamma$ agree with the analytic computation of the largest Lyapunov exponent, where the scaling constant $L=50$ in Fig.~\ref{DecayExponents}. The crossover of the largest Lyapunov exponent indicates the existence of a threshold corresponding to the transition from weak chaos to strong chaos.

In the weak chaos regime ($\beta \to 0$), the harmonic coupling plays the dominant role and leads to phonons and their interactions.

In the transition regime, a moderate nonlinearity allows the emergence of solitary waves. The phase space of the Hamiltonian system is composed of chaotic trajectories intermingled with KAM tori. The multiple-peak structure of the lifetime of the solitary wave observed in Fig.~\ref{Lifetime} is found in this transition regime and can be explained by the transiently quasi-regular motions in phase space induced by moderate stochasticity, where KAM invariant tori are dominant. The long-transient solitary wave states dynamically correspond to these KAM tori, and the lifetime of the solitary wave actually implies the relative stability of the corresponding KAM torus. These results may provide useful hints in understanding the stability of KAM tori in high-dimensional Hamiltonian systems.

In the strong chaos regime, the nonlinear coupling plays the main role and leads to the dominance of solitary waves. We discuss below the relation between the macroscopic heat phenomena on low-dimensional nonlinear lattices and our microscopic results, which is the initial motivation of our present work.

The anomalous thermal conductivity of low-dimensional nonlinear lattices is not completely well theoretically understood now. Based on the effective phonon theory and a conjecture that the mean-free-path of the effective phonons is inversely proportional to the dimensionless nonlinearity $\xi$ as a ratio between the average of nonlinear potential energy and the total potential energy, temperature dependence of thermal conductivity in the FPU-$\beta$ model is well explained~\cite{LiEPL2007}.

We can directly estimate the mean-free-path of the solitary waves and the interesting thing is that our estimation is the same as the conjecture of the mean-free-path for the effective phonons at the high temperature regime~\cite{LiEPL2007}. This indicates the similarity between the solitary waves and the effective phonons at the high temperature regime. The process of our estimation is presented as follows. The large energy scaling of the solitary wave velocity with the energy as $v \propto E^{1/4}$ is well know~\cite{Aoki2001}. Since changing the nonlinear parameter $\beta$ is equivalent to changing the energy density, we have the scaling of the solitary wave velocity with the nonlinear parameter $\beta$, $v \propto \beta^{1/4}$. Microscopically, we show in our work that the motion of the solitary wave on the finite FPU-$\beta$ lattice may give rise to phonon tails. The interaction between the solitary wave and the phonon waves leads to the instability of the solitary wave, especially in the strong chaos regime. The existence of thermal baths on the boundaries, which are usually adopted in studies of thermal conductions of low-dimensional lattices, may drastically enhance the excitations of more phonons and the instability of the solitary wave. According to the exponential decay law of the solitary wave during the final loss of stability, the relaxation time of the solitary wave can be estimated as $\tau \propto 1/\gamma$. For the decay rate $\gamma$ obeys the scaling law $\gamma \propto \beta^{1/4}$ in the strong chaos regime, the mean-free-path of the solitary wave can be estimated to be $\l=v \tau \propto 1$.

\section{Concluding remarks}

In conclusion, in this Letter we extensively explored the spatiotemporal propagation behavior of a momentum excitation traveling along the FPU-$\beta$ lattice with finite size for different nonlinear strengths and different boundary conditions. For a moderate nonlinearity, the solitary wave can coherently propagate along the lattice for a long time and then decays rapidly after this transient due to the final dominance of phonons. The lifetime of the long-transient propagation state is sensitive to the value of the anharmonic parameter $\beta$ and exhibits a fractal dependence on $\beta$ for the case of open boundary condition.

The energy of the solitary wave is found to decay exponentially during the final loss of stability, which is an intrinsic property of the lattice independence of the type of boundary conditions and the parameters of the system. The decay rate $\gamma$ of the solitary wave as the function of $\beta$ exhibits two different scaling laws, which is consistent with the scenario predicted in the largest Lyapunov exponent of the FPU-$\beta$ model. Therefore the loss of stability of the solitary wave is a manifestation of dynamical instability of orbits in the Hamiltonian system. We found that the appearance of the hierarchical multiple-peak structure of the lifetime of the long-transient propagation state is in cases of $\beta$ in the transition regime from weak chaos to strong chaos. Therefore this interesting result can be well explained by the residuals of KAM tori in the phase space of the Hamiltonian system in the parameter regime of $\beta$ with the moderate stochasticity, which induce the "stickness" effect despite of their instability in high-dimensional Hamiltonian cases. In fact, the multiple-peak of the lifetime is closely related to the structure of these unstable KAM tori in phase space.

To solve the debate about the energy carriers responsible for the heat conduction in the FPU-$\beta$ lattice, the sound velocity of energy transfer was measured to examine the properties of the energy carriers, by using both nonequilibrium and equilibrium approaches. Nevertheless, the uncertainty of the computational data is too large to distinguish between the two predictions based on soliton theory and effective phonon theory. Our discussion of mean-free-path of the solitary waves may be helpful to understand the origin of anomalous conductivity and the debate about the energy carriers in the FPU-$\beta$ lattice~\cite{Zhang2000,Zhang2001,Aoki2001,Zhao2006,Li2006,Li2010}.

\acknowledgments
Project supported by the National Natural Science Foundation of China (Grant No. 11075016), the Fundamental Research Funds for the Central Universities of China (Grant No. 201001), and the Research Fund for the Doctoral Program of Higher Education of China (Grant No. 20100003110007).

\bibliography{References}

\begin{thebibliography}{39}
\expandafter\ifx\csname natexlab\endcsname\relax\def\natexlab#1{#1}\fi
\expandafter\ifx\csname bibnamefont\endcsname\relax
  \def\bibnamefont#1{#1}\fi
\expandafter\ifx\csname bibfnamefont\endcsname\relax
  \def\bibfnamefont#1{#1}\fi
\expandafter\ifx\csname citenamefont\endcsname\relax
  \def\citenamefont#1{#1}\fi
\expandafter\ifx\csname url\endcsname\relax
  \def\url#1{\texttt{#1}}\fi
\expandafter\ifx\csname urlprefix\endcsname\relax\def\urlprefix{URL }\fi
\providecommand{\bibinfo}[2]{#2}
\providecommand{\eprint}[2][]{\url{#2}}

\bibitem[{\citenamefont{Lepri et~al.}(1997)\citenamefont{Lepri, Livi, and
  Politi}}]{Lepri1997}
\bibinfo{author}{\bibfnamefont{S.}~\bibnamefont{Lepri}},
  \bibinfo{author}{\bibfnamefont{R.}~\bibnamefont{Livi}}, \bibnamefont{and}
  \bibinfo{author}{\bibfnamefont{A.}~\bibnamefont{Politi}},
  \bibinfo{journal}{Phys. Rev. Lett.} \textbf{\bibinfo{volume}{78}},
  \bibinfo{pages}{1896} (\bibinfo{year}{1997}).

\bibitem[{\citenamefont{Aoki and Kusnezov}(2001)}]{Aoki2001}
\bibinfo{author}{\bibfnamefont{K.}~\bibnamefont{Aoki}} \bibnamefont{and}
  \bibinfo{author}{\bibfnamefont{D.}~\bibnamefont{Kusnezov}},
  \bibinfo{journal}{Phys. Rev. Lett.} \textbf{\bibinfo{volume}{86}},
  \bibinfo{pages}{4029} (\bibinfo{year}{2001}).

\bibitem[{\citenamefont{Lepri et~al.}(2003)\citenamefont{Lepri, Livi, and
  Politi}}]{Lepri2003}
\bibinfo{author}{\bibfnamefont{S.}~\bibnamefont{Lepri}},
  \bibinfo{author}{\bibfnamefont{R.}~\bibnamefont{Livi}}, \bibnamefont{and}
  \bibinfo{author}{\bibfnamefont{A.}~\bibnamefont{Politi}},
  \bibinfo{journal}{Phys. Rep.} \textbf{\bibinfo{volume}{377}},
  \bibinfo{pages}{1} (\bibinfo{year}{2003}).

\bibitem[{\citenamefont{Dhar}(2008)}]{Dhar2008c}
\bibinfo{author}{\bibfnamefont{A.}~\bibnamefont{Dhar}}, \bibinfo{journal}{Adv.
  Phys.} \textbf{\bibinfo{volume}{57}}, \bibinfo{pages}{457}
  (\bibinfo{year}{2008}).

\bibitem[{\citenamefont{Li et~al.}(2012)\citenamefont{Li, Ren, Wang, Zhang,
  H\"anggi, and Li}}]{Li2012}
\bibinfo{author}{\bibfnamefont{N.}~\bibnamefont{Li}},
  \bibinfo{author}{\bibfnamefont{J.}~\bibnamefont{Ren}},
  \bibinfo{author}{\bibfnamefont{L.}~\bibnamefont{Wang}},
  \bibinfo{author}{\bibfnamefont{G.}~\bibnamefont{Zhang}},
  \bibinfo{author}{\bibfnamefont{P.}~\bibnamefont{H\"anggi}}, \bibnamefont{and}
  \bibinfo{author}{\bibfnamefont{B.}~\bibnamefont{Li}}, \bibinfo{journal}{Rev.
  Mod. Phys.} \textbf{\bibinfo{volume}{84}}, \bibinfo{pages}{1045}
  (\bibinfo{year}{2012}).

\bibitem[{\citenamefont{Terraneo et~al.}(2002)\citenamefont{Terraneo, Peyrard,
  and Casati}}]{Terraneo2002}
\bibinfo{author}{\bibfnamefont{M.}~\bibnamefont{Terraneo}},
  \bibinfo{author}{\bibfnamefont{M.}~\bibnamefont{Peyrard}}, \bibnamefont{and}
  \bibinfo{author}{\bibfnamefont{G.}~\bibnamefont{Casati}},
  \bibinfo{journal}{Phys. Rev. Lett.} \textbf{\bibinfo{volume}{88}},
  \bibinfo{pages}{094302} (\bibinfo{year}{2002}).

\bibitem[{\citenamefont{Li et~al.}(2004)\citenamefont{Li, Wang, and
  Casati}}]{Li2004}
\bibinfo{author}{\bibfnamefont{B.}~\bibnamefont{Li}},
  \bibinfo{author}{\bibfnamefont{L.}~\bibnamefont{Wang}}, \bibnamefont{and}
  \bibinfo{author}{\bibfnamefont{G.}~\bibnamefont{Casati}},
  \bibinfo{journal}{Phys. Rev. Lett.} \textbf{\bibinfo{volume}{93}},
  \bibinfo{pages}{184301} (\bibinfo{year}{2004}).

\bibitem[{\citenamefont{Chang et~al.}(2007)\citenamefont{Chang, Okawa, Garcia,
  Majumdar, and Zettl}}]{Chang2007a}
\bibinfo{author}{\bibfnamefont{C.~W.} \bibnamefont{Chang}},
  \bibinfo{author}{\bibfnamefont{D.}~\bibnamefont{Okawa}},
  \bibinfo{author}{\bibfnamefont{H.}~\bibnamefont{Garcia}},
  \bibinfo{author}{\bibfnamefont{A.}~\bibnamefont{Majumdar}}, \bibnamefont{and}
  \bibinfo{author}{\bibfnamefont{A.}~\bibnamefont{Zettl}},
  \bibinfo{journal}{Phys. Rev. Lett.} \textbf{\bibinfo{volume}{99}},
  \bibinfo{pages}{045901} (\bibinfo{year}{2007}).

\bibitem[{\citenamefont{Wang and Li}(2007)}]{Wang2007}
\bibinfo{author}{\bibfnamefont{L.}~\bibnamefont{Wang}} \bibnamefont{and}
  \bibinfo{author}{\bibfnamefont{B.}~\bibnamefont{Li}}, \bibinfo{journal}{Phys.
  Rev. Lett.} \textbf{\bibinfo{volume}{99}}, \bibinfo{pages}{177208}
  (\bibinfo{year}{2007}).

\bibitem[{\citenamefont{Wang and Li}(2008{\natexlab{a}})}]{Wang2008a}
\bibinfo{author}{\bibfnamefont{L.}~\bibnamefont{Wang}} \bibnamefont{and}
  \bibinfo{author}{\bibfnamefont{B.}~\bibnamefont{Li}}, \bibinfo{journal}{Phys.
  Rev. Lett.} \textbf{\bibinfo{volume}{101}}, \bibinfo{pages}{267203}
  (\bibinfo{year}{2008}{\natexlab{a}}).

\bibitem[{\citenamefont{Wang and Li}(2008{\natexlab{b}})}]{Wang2008}
\bibinfo{author}{\bibfnamefont{L.}~\bibnamefont{Wang}} \bibnamefont{and}
  \bibinfo{author}{\bibfnamefont{B.}~\bibnamefont{Li}}, \bibinfo{journal}{Phys.
  World} \textbf{\bibinfo{volume}{21}}, \bibinfo{pages}{27}
  (\bibinfo{year}{2008}{\natexlab{b}}).

\bibitem[{\citenamefont{Chang et~al.}(2006)\citenamefont{Chang, Okawa,
  Majumdar, and Zettl}}]{Chang2006a}
\bibinfo{author}{\bibfnamefont{C.~W.} \bibnamefont{Chang}},
  \bibinfo{author}{\bibfnamefont{D.}~\bibnamefont{Okawa}},
  \bibinfo{author}{\bibfnamefont{A.}~\bibnamefont{Majumdar}}, \bibnamefont{and}
  \bibinfo{author}{\bibfnamefont{A.}~\bibnamefont{Zettl}},
  \bibinfo{journal}{Science} \textbf{\bibinfo{volume}{314}},
  \bibinfo{pages}{1121} (\bibinfo{year}{2006}).

\bibitem[{\citenamefont{Kobayashi et~al.}(2009)\citenamefont{Kobayashi,
  Teraoka, and Terasaki}}]{Kobayashi2009}
\bibinfo{author}{\bibfnamefont{W.}~\bibnamefont{Kobayashi}},
  \bibinfo{author}{\bibfnamefont{Y.}~\bibnamefont{Teraoka}}, \bibnamefont{and}
  \bibinfo{author}{\bibfnamefont{I.}~\bibnamefont{Terasaki}},
  \bibinfo{journal}{Appl. Phys. Lett.} \textbf{\bibinfo{volume}{95}},
  \bibinfo{pages}{171905} (\bibinfo{year}{2009}).

\bibitem[{\citenamefont{Ren and Li}(2010)}]{Ren2010}
\bibinfo{author}{\bibfnamefont{J.}~\bibnamefont{Ren}} \bibnamefont{and}
  \bibinfo{author}{\bibfnamefont{B.}~\bibnamefont{Li}}, \bibinfo{journal}{Phys.
  Rev. E} \textbf{\bibinfo{volume}{81}}, \bibinfo{pages}{021111}
  (\bibinfo{year}{2010}).

\bibitem[{\citenamefont{Zavt et~al.}(1993)\citenamefont{Zavt, Wagner, and
  L\"utze}}]{Zavt1993}
\bibinfo{author}{\bibfnamefont{G.~S.} \bibnamefont{Zavt}},
  \bibinfo{author}{\bibfnamefont{M.}~\bibnamefont{Wagner}}, \bibnamefont{and}
  \bibinfo{author}{\bibfnamefont{A.}~\bibnamefont{L\"utze}},
  \bibinfo{journal}{Phys. Rev. E} \textbf{\bibinfo{volume}{47}},
  \bibinfo{pages}{4108} (\bibinfo{year}{1993}).

\bibitem[{\citenamefont{Sarmiento et~al.}(1999)\citenamefont{Sarmiento,
  Reigada, Romero, and Lindenberg}}]{Sarmiento1999}
\bibinfo{author}{\bibfnamefont{A.}~\bibnamefont{Sarmiento}},
  \bibinfo{author}{\bibfnamefont{R.}~\bibnamefont{Reigada}},
  \bibinfo{author}{\bibfnamefont{A.~H.} \bibnamefont{Romero}},
  \bibnamefont{and}
  \bibinfo{author}{\bibfnamefont{K.}~\bibnamefont{Lindenberg}},
  \bibinfo{journal}{Phys. Rev. E} \textbf{\bibinfo{volume}{60}},
  \bibinfo{pages}{5317} (\bibinfo{year}{1999}).

\bibitem[{\citenamefont{Rosas and Lindenberg}(2004)}]{Rosas2004}
\bibinfo{author}{\bibfnamefont{A.}~\bibnamefont{Rosas}} \bibnamefont{and}
  \bibinfo{author}{\bibfnamefont{K.}~\bibnamefont{Lindenberg}},
  \bibinfo{journal}{Phys. Rev. E} \textbf{\bibinfo{volume}{69}},
  \bibinfo{pages}{016615} (\bibinfo{year}{2004}).

\bibitem[{\citenamefont{Fermi et~al.}(1955)\citenamefont{Fermi, Pasta, and
  Ulam}}]{Fermi1955}
\bibinfo{author}{\bibfnamefont{E.}~\bibnamefont{Fermi}},
  \bibinfo{author}{\bibfnamefont{J.}~\bibnamefont{Pasta}}, \bibnamefont{and}
  \bibinfo{author}{\bibfnamefont{S.}~\bibnamefont{Ulam}}, \bibinfo{journal}{Los
  Alamos Document No. LA-1940}  (\bibinfo{year}{1955}).

\bibitem[{\citenamefont{Zabusky and Kruskal}(1965)}]{Zabusky1965}
\bibinfo{author}{\bibfnamefont{N.~J.} \bibnamefont{Zabusky}} \bibnamefont{and}
  \bibinfo{author}{\bibfnamefont{M.~D.} \bibnamefont{Kruskal}},
  \bibinfo{journal}{Phys. Rev. Lett.} \textbf{\bibinfo{volume}{15}},
  \bibinfo{pages}{240} (\bibinfo{year}{1965}).

\bibitem[{\citenamefont{Sievers and Takeno}(1988)}]{Sievers1988}
\bibinfo{author}{\bibfnamefont{A.~J.} \bibnamefont{Sievers}} \bibnamefont{and}
  \bibinfo{author}{\bibfnamefont{S.}~\bibnamefont{Takeno}},
  \bibinfo{journal}{Phys. Rev. Lett.} \textbf{\bibinfo{volume}{61}},
  \bibinfo{pages}{970} (\bibinfo{year}{1988}).

\bibitem[{\citenamefont{Flach et~al.}(2005)\citenamefont{Flach, Ivanchenko, and
  Kanakov}}]{Flach2005}
\bibinfo{author}{\bibfnamefont{S.}~\bibnamefont{Flach}},
  \bibinfo{author}{\bibfnamefont{M.~V.} \bibnamefont{Ivanchenko}},
  \bibnamefont{and} \bibinfo{author}{\bibfnamefont{O.~I.}
  \bibnamefont{Kanakov}}, \bibinfo{journal}{Phys. Rev. Lett.}
  \textbf{\bibinfo{volume}{95}}, \bibinfo{pages}{064102}
  (\bibinfo{year}{2005}).

\bibitem[{\citenamefont{Villain and Lewenstein}(2000)}]{Villain2000}
\bibinfo{author}{\bibfnamefont{P.}~\bibnamefont{Villain}} \bibnamefont{and}
  \bibinfo{author}{\bibfnamefont{M.}~\bibnamefont{Lewenstein}},
  \bibinfo{journal}{Phys. Rev. A} \textbf{\bibinfo{volume}{62}},
  \bibinfo{pages}{043601} (\bibinfo{year}{2000}).

\bibitem[{\citenamefont{Miloshevich et~al.}(2009)\citenamefont{Miloshevich,
  Khomeriki, and Ruffo}}]{Miloshevich2009}
\bibinfo{author}{\bibfnamefont{G.}~\bibnamefont{Miloshevich}},
  \bibinfo{author}{\bibfnamefont{R.}~\bibnamefont{Khomeriki}},
  \bibnamefont{and} \bibinfo{author}{\bibfnamefont{S.}~\bibnamefont{Ruffo}},
  \bibinfo{journal}{Phys. Rev. Lett.} \textbf{\bibinfo{volume}{102}},
  \bibinfo{pages}{020602} (\bibinfo{year}{2009}).

\bibitem[{\citenamefont{Li et~al.}(2005)\citenamefont{Li, Lan, and
  Wang}}]{Li2005a}
\bibinfo{author}{\bibfnamefont{B.}~\bibnamefont{Li}},
  \bibinfo{author}{\bibfnamefont{J.~H.} \bibnamefont{Lan}}, \bibnamefont{and}
  \bibinfo{author}{\bibfnamefont{L.}~\bibnamefont{Wang}},
  \bibinfo{journal}{Phys. Rev. Lett.} \textbf{\bibinfo{volume}{95}},
  \bibinfo{pages}{104302} (\bibinfo{year}{2005}).

\bibitem[{\citenamefont{Mai et~al.}(2007)\citenamefont{Mai, Dhar, and
  Narayan}}]{Mai2007}
\bibinfo{author}{\bibfnamefont{T.}~\bibnamefont{Mai}},
  \bibinfo{author}{\bibfnamefont{A.}~\bibnamefont{Dhar}}, \bibnamefont{and}
  \bibinfo{author}{\bibfnamefont{O.}~\bibnamefont{Narayan}},
  \bibinfo{journal}{Phys. Rev. Lett.} \textbf{\bibinfo{volume}{98}},
  \bibinfo{pages}{184301} (\bibinfo{year}{2007}).

\bibitem[{\citenamefont{Dhar and Saito}(2008)}]{Dhar2008b}
\bibinfo{author}{\bibfnamefont{A.}~\bibnamefont{Dhar}} \bibnamefont{and}
  \bibinfo{author}{\bibfnamefont{K.}~\bibnamefont{Saito}},
  \bibinfo{journal}{Phys. Rev. E} \textbf{\bibinfo{volume}{78}},
  \bibinfo{pages}{061136} (\bibinfo{year}{2008}).

\bibitem[{\citenamefont{Li et~al.}(2010)\citenamefont{Li, Li, and
  Flach}}]{Li2010}
\bibinfo{author}{\bibfnamefont{N.}~\bibnamefont{Li}},
  \bibinfo{author}{\bibfnamefont{B.}~\bibnamefont{Li}}, \bibnamefont{and}
  \bibinfo{author}{\bibfnamefont{S.}~\bibnamefont{Flach}},
  \bibinfo{journal}{Phys. Rev. Lett.} \textbf{\bibinfo{volume}{105}},
  \bibinfo{pages}{054102} (\bibinfo{year}{2010}).

\bibitem[{\citenamefont{Antonopoulos and Bountis}(2006)}]{Antonopoulos2006}
\bibinfo{author}{\bibfnamefont{C.}~\bibnamefont{Antonopoulos}}
  \bibnamefont{and} \bibinfo{author}{\bibfnamefont{T.}~\bibnamefont{Bountis}},
  \bibinfo{journal}{Phys. Rev. E} \textbf{\bibinfo{volume}{73}},
  \bibinfo{pages}{056206} (\bibinfo{year}{2006}).

\bibitem[{\citenamefont{Wattis}(1993)}]{Wattis1993}
\bibinfo{author}{\bibfnamefont{J.~A.~D.} \bibnamefont{Wattis}},
  \bibinfo{journal}{J. Phys. A} \textbf{\bibinfo{volume}{26}},
  \bibinfo{pages}{1193} (\bibinfo{year}{1993}).

\bibitem[{\citenamefont{Friesecke and Wattis}(1994)}]{Friesecke1994}
\bibinfo{author}{\bibfnamefont{G.}~\bibnamefont{Friesecke}} \bibnamefont{and}
  \bibinfo{author}{\bibfnamefont{J.}~\bibnamefont{Wattis}},
  \bibinfo{journal}{Commun. Math. Phys.} \textbf{\bibinfo{volume}{161}},
  \bibinfo{pages}{391} (\bibinfo{year}{1994}).

\bibitem[{\citenamefont{Zhang et~al.}(2000)\citenamefont{Zhang, Isbister, and
  Evans}}]{Zhang2000}
\bibinfo{author}{\bibfnamefont{F.}~\bibnamefont{Zhang}},
  \bibinfo{author}{\bibfnamefont{D.~J.} \bibnamefont{Isbister}},
  \bibnamefont{and} \bibinfo{author}{\bibfnamefont{D.~J.} \bibnamefont{Evans}},
  \bibinfo{journal}{Phys. Rev. E} \textbf{\bibinfo{volume}{61}},
  \bibinfo{pages}{3541} (\bibinfo{year}{2000}).

\bibitem[{\citenamefont{Zhang et~al.}(2001)\citenamefont{Zhang, Isbister, and
  Evans}}]{Zhang2001}
\bibinfo{author}{\bibfnamefont{F.}~\bibnamefont{Zhang}},
  \bibinfo{author}{\bibfnamefont{D.~J.} \bibnamefont{Isbister}},
  \bibnamefont{and} \bibinfo{author}{\bibfnamefont{D.~J.} \bibnamefont{Evans}},
  \bibinfo{journal}{Phys. Rev. E} \textbf{\bibinfo{volume}{64}},
  \bibinfo{pages}{021102} (\bibinfo{year}{2001}).

\bibitem[{\citenamefont{Forrest and Jr}(1979)}]{Forrest1979}
\bibinfo{author}{\bibfnamefont{S.~R.} \bibnamefont{Forrest}} \bibnamefont{and}
  \bibinfo{author}{\bibfnamefont{T.~A.~W.} \bibnamefont{Jr}},
  \bibinfo{journal}{J. Phys. A} \textbf{\bibinfo{volume}{12}},
  \bibinfo{pages}{L109} (\bibinfo{year}{1979}).

\bibitem[{\citenamefont{Gallavotti}(2008)}]{Gallavotti2008}
\bibinfo{editor}{\bibfnamefont{G.}~\bibnamefont{Gallavotti}}, ed.,
  \emph{\bibinfo{title}{The Fermi-Pasta-Ulam problem}}
  (\bibinfo{publisher}{Springer Verlag}, \bibinfo{year}{2008}).

\bibitem[{\citenamefont{Casetti et~al.}(1995)\citenamefont{Casetti, Livi, and
  Pettini}}]{Casetti1995}
\bibinfo{author}{\bibfnamefont{L.}~\bibnamefont{Casetti}},
  \bibinfo{author}{\bibfnamefont{R.}~\bibnamefont{Livi}}, \bibnamefont{and}
  \bibinfo{author}{\bibfnamefont{M.}~\bibnamefont{Pettini}},
  \bibinfo{journal}{Phys. Rev. Lett.} \textbf{\bibinfo{volume}{74}},
  \bibinfo{pages}{375} (\bibinfo{year}{1995}).

\bibitem[{\citenamefont{Casetti et~al.}(1996)\citenamefont{Casetti, Clementi,
  and Pettini}}]{Casetti1996}
\bibinfo{author}{\bibfnamefont{L.}~\bibnamefont{Casetti}},
  \bibinfo{author}{\bibfnamefont{C.}~\bibnamefont{Clementi}}, \bibnamefont{and}
  \bibinfo{author}{\bibfnamefont{M.}~\bibnamefont{Pettini}},
  \bibinfo{journal}{Phys. Rev. E} \textbf{\bibinfo{volume}{54}},
  \bibinfo{pages}{5969} (\bibinfo{year}{1996}).

\bibitem[{\citenamefont{Li and Li}(2007)}]{LiEPL2007}
\bibinfo{author}{\bibfnamefont{N.}~\bibnamefont{Li}} \bibnamefont{and}
  \bibinfo{author}{\bibfnamefont{B.}~\bibnamefont{Li}},
  \bibinfo{journal}{Europhys. Lett.} \textbf{\bibinfo{volume}{78}},
  \bibinfo{pages}{34001} (\bibinfo{year}{2007}).

\bibitem[{\citenamefont{Zhao}(2006)}]{Zhao2006}
\bibinfo{author}{\bibfnamefont{H.}~\bibnamefont{Zhao}}, \bibinfo{journal}{Phys.
  Rev. Lett.} \textbf{\bibinfo{volume}{96}}, \bibinfo{pages}{140602}
  (\bibinfo{year}{2006}).

\bibitem[{\citenamefont{Li et~al.}(2006)\citenamefont{Li, Tong, and
  Li}}]{Li2006}
\bibinfo{author}{\bibfnamefont{N.}~\bibnamefont{Li}},
  \bibinfo{author}{\bibfnamefont{P.}~\bibnamefont{Tong}}, \bibnamefont{and}
  \bibinfo{author}{\bibfnamefont{B.}~\bibnamefont{Li}},
  \bibinfo{journal}{Europhys. Lett.} \textbf{\bibinfo{volume}{75}},
  \bibinfo{pages}{49} (\bibinfo{year}{2006}).

\end{thebibliography}

\end{document}